\documentclass[a4paper,english,final]{IEEEtran}
\usepackage{amsmath}
\usepackage{amssymb}
\usepackage{graphicx}
\usepackage{setspace}
\usepackage{blindtext, graphicx}
\usepackage{amsmath}
\usepackage{amssymb}
\usepackage{amsfonts}
\usepackage{float}
\usepackage{cuted}
\usepackage{amsthm}
\usepackage{xcolor}
\newtheoremstyle{case}{}{}{}{}{}{:}{ }{}

\usepackage{stfloats}

\usepackage[ruled,vlined]{algorithm2e}[]
\makeatletter
\IEEEoverridecommandlockouts
\usepackage{cite}
\usepackage{amsfonts}\usepackage{algorithmic}
\usepackage{textcomp}

\def\BibTeX{{\rm B\kern-.05em{\sc i\kern-.025em b}\kern-.08em
    T\kern-.1667em\lower.7ex\hbox{E}\kern-.125emX}}

\usepackage{bm}
\usepackage{cite}\usepackage{amsthm}
\usepackage{epsfig}
\usepackage{latexsym}
\usepackage{wasysym}
\usepackage{placeins}
\usepackage{multicol}
\usepackage{here}
\usepackage{subfigure}

\DeclareMathAlphabet{\mathbit}{OML}{cmr}{bx}{it}
\DeclareMathAlphabet{\mathsf}{OT1}{cmss}{m}{n}
\DeclareMathAlphabet{\mathTXf}{OT1}{cmss}{bx}{it}

\usepackage[acronym]{glossaries}
\newcommand{\newac}{\newacronym}

\makeglossaries

\newac{los}{LOS}{line-of-sight}
\newac{uav}{UAV}{unmanned aerial vehicle}

\usepackage{verbatim}

\makeatother

\begin{document}
\title{UAV-aided RF Mapping for Sensing and Connectivity in Wireless Networks
}

\author{
  \IEEEauthorblockN{David Gesbert, Omid Esrafilian, Junting Chen, Rajeev Gangula, Urbashi Mitra }
    \thanks{David Gesbert, Omid Esrafilian, and Rajeev Gangula are with the Communication
Systems Department, EURECOM. Emails: \{gesbert, esrafili, gangula \}@eurecom.fr.\newline Junting Chen is with School of Science and Engineering and Future Network of Intelligence Institute, The Chinese University of Hong Kong, Shenzhen. Email: juntingc@cuhk.edu.cn.\newline Urbashi Mitra is with Ming Hsieh Department of Electrical Engineering, University of Southern California. Email: ubli@usc.edu.}
}

\maketitle

\begin{abstract}
 The use of unmanned aerial vehicles (UAV) as flying radio access network (RAN) nodes offers a promising complement to traditional fixed terrestrial deployments. More recently yet still in the context of wireless networks,  drones have also been envisioned for use as radio frequency (RF) sensing and localization devices.  
In both cases, the advantage of using UAVs lies in their ability to navigate themselves freely in 3D and in a timely manner to locations of space where the obtained network throughput or sensing performance is optimal. In practice, the selection of a proper location or trajectory for the UAV very much depends on local terrain features, including the position of surrounding radio obstacles. Hence, the robot must be able to {\em map} the features of its radio environment as it performs its data communication or sensing services. The challenges related to this task, referred here as {\em radio mapping},
are discussed in this paper. Its promises related to efficient trajectory design for autonomous radio-aware UAVs are highlighted, along with algorithm solutions. The advantages induced by radio-mapping in terms  of connectivity, sensing, and localization performance are illustrated.

\end{abstract}


\section{UAV-aided networks and placement problems}



\begin{figure}[ht!]
\centering \includegraphics[width=0.8\columnwidth]{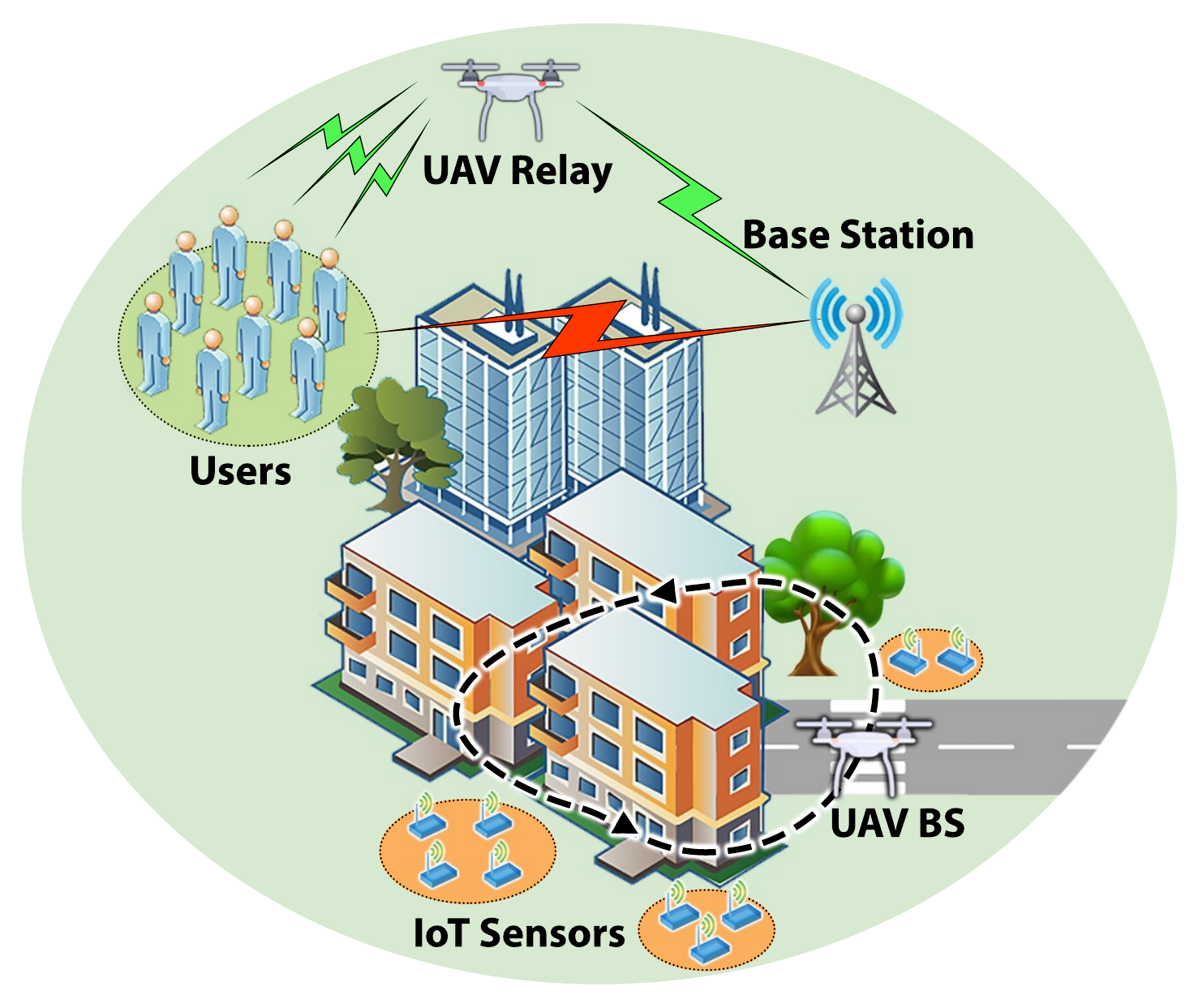} \caption[]{Illustration of some promising flying RANs use cases: Drone-as-a-relay
for ground user connectivity (DaaR), Drone-as-a-base station for IoT
data harvesting (DaaB).}
\label{IA-JP}
\end{figure}

The exploitation of drones, aka UAVs, within
the future 6G wireless cellular communication networks has recently gained
significant attention. Several scenarios have been articulated in
the literature which we can be categorize
as Drone-as-a-Terminal (DaaT), Drone-as-a-Relay (DaaR), Drone-as-a-Base
station (DaaB), and Drone-as-a-Sensor (DaaS) scenarios, respectively. In the DaaT scenario, applications range from delivery to monitoring and surveillance, and wireless networks play a vital role to carry UAV control (possibly video-based) and command data.
In contrast, the DaaR and DaaB frameworks view the UAV as a
piece of the radio access network (RAN) infrastructure, as shown in
Fig. \ref{IA-JP}. The UAV acts as a flying base station (BS) which can,
for example, harvest data sent from ground nodes. The UAV can also be 
a flying real-time relay to extend coverage from a ground-based fixed BS.
A promising feature of both DaaB and DaaR scenarios is to allow
a flexible deployment of radio resources {\em when} and {\em
where} they are most needed. Use cases range from disaster recovery
scenarios, servicing of temporary cultural/sporting events, road traffic
assistance, hot-spots coverage, and Internet-of-Things (IoT) data
harvesting (smart city, agriculture, ...) \cite{wu2021comprehensive}.
In DaaS applications, the UAV acts as a flying sensor collecting (radio) data for radio frequency (RF) sensing and localization purposes, which are important and novel use cases for 6G. 

 While research challenges dealing with radio-aided UAVs and UAV-aided radio networks are plenty,
the problem of how UAVs can best (self-) navigate the radio environment to render the best possible communication or sensing services remains perhaps the most critical and fascinating issue \cite{wu2021comprehensive}.

In order to offer much needed performance guarantees, the trajectory design algorithms must
be adaptive to context parameters, such as ground radio node locations,
the traffic distribution, the quality of service (QoS) or sensing requirements,
and the propagation conditions shaped by the radio obstacles. Ideally, the algorithm operates in
an autonomous fashion,  either on-board
the drone or in a ground-based computing unit that is connected to
the drone. From an algorithmic perspective, it is important to distinguish
between the (i) {\em static} placement problem from (ii) {\em
path} or trajectory planning. Static placement involves finding a
single good 3D location for the UAV, from where to provide connectivity
to not-too distant ground nodes or sense the environment. While the solution may be updated
when large-scale system parameters vary, such as traffic or ground
user location distributions, the UAV location is otherwise stable
and can benefit from energy-saving mechanisms, such as  the ability to exploit nearby {\em resting spots} \cite{gangula18b}.
 In some scenarios, however,
there is interest in flying along an optimal path. For DaaB scenarios, a pattern that brings
the UAV closer in turn to each ground node will  improve the average throughput over a static deployment
\cite{ZenZha:J17} or reduce energy expenditure at the nodes in an IoT setting \cite{Esra19}. Path optimization
may also take into account specific kinematic energy consumption
models, obstacle avoidance, as well as realistic robot
dynamics, leading to
a potentially complex mixed communication-robotics optimization framework \cite{wu2021comprehensive}.

Regardless of the static or dynamic nature of the placement strategy,
the algorithms usually operate on the basis of an array of information
which may include ground node GPS location information, per-node data
traffic requirements and, importantly, terrain-dependent propagation
data allowing the reliable {\it prediction of radio signal strengths}. While
such data may be collected via the network beforehand allowing placement
to be optimized before the actual UAV flight, part or all of the information
may also have to be {\em discovered} or {\em learned} by the
UAV while {\em in flight} to its destination, implying some degree
of {\em online} optimization. The choice between
the offline and online cases gives rise to an interesting trade-off
between flight efficiency and adaptability vis-\`a-vis a priori unknown
deployment settings.

In the DaaB and DaaR settings, the premium offered  over  fixed cellular deployment
 essentially lies in the ability to bring the RAN closer
to the user so as to increase the radio channel quality. In the DaaS case, the designed trajectory aims at optimally enriching the set of measurements collected along the path in order to accelerate sensing performance.

In all these cases, the
influence of channel models in the placement solution
is critical. The assumption of Line-of-sight (LoS) channels or the use of simple statistical
blocking models (i.e. modeling the LoS probability) has proved an
excellent way to derive early insights into the problem  \cite{HouSitLar:J14,wu2021comprehensive}.
Unfortunately, the probabilistic nature of such approaches
limits our ability to {\em guarantee} performance in an actual
UAV deployment. For example, a statistically optimized placement of a flying BS
might suggest a location which one eventually discovers to be
severely affected by local blockage (e.g. unforeseen presence
of a tall building) forcing the drone to recompute a sub-optimal path.

 
As an alternative to probabilistic channel models, trajectory design can instead exploit channel prediction based on local terrain features. To this end, the UAV can map out its environment or exploit readily available maps in order to navigate through the radio environment with enhanced reliability. As such, radio maps or 3D physical maps play an essential role in optimal UAV-based connectivity and sensing. The problem of estimating such maps from data collected at the UAV and suitable channel models is looked at in the next sections, followed by methods for placing the UAV optimally on the basis of the maps. The problem of optimally placing the UAV for communication or sensing purposes based on map data  unfortunately comes with its own challenges. One of them is referred to as the "rich data dilemma": On the one hand, radio or 3D maps are rich in useful contextual information, revealing the position of each radio obstacle. On the other, radio blockage is often irregular and causes a lack of structure in the map, breaking differentiability in the placement optimization problem.  As shown in this paper, a possible solution to this issue can be found in the concept of {\it map compression} which attempts to retain the "local knowledge" feature of maps while lending the differentiability aspect to the placement problem, making it amenable to standard optimization.
Finally, this paper discusses  the optimization of the flight path toward optimal sensing through  robotics-inspired active learning principles. Section VI briefly presents practical prototype realizations.

\section{UAV-to-ground segmented channel models}


Air-to-ground channel modeling plays an essential role in our problem. The quality of channel models lies in their ability
to correctly predict power attenuation as a function of distance,
frequency, as well as to characterize seemingly random blockage. While
some models encompass multi-path fading, the placement optimization
time scale is usually much longer than the fast fading coherence time,
hence fast fading can be averaged out in a first approximation, leaving the design with the task of  modelling the path loss or received signal strength indicator (RSSI).
In contrast to terrestrial radio pathloss models, air-to-ground channels between low altitude flying radios and ground users are known to exhibit a {\it segmented} nature \cite{HouSitLar:J14} that depends on the relative locations of UAV and user. 
 In such models the path loss vs. distance follows the classical exponential decay rule where the exponent $\alpha_s$ is selected from a discrete set of values $(s=1,..,K)$ and
where the so-called
{\em segment value} $s$ reflects the degree of link's obstruction
and its strong dependence on local terrain.
As an example, we can differentiate the LoS and non-LoS (NLoS) nature of air-to-ground links by using a segmented model with $K=2$, $\alpha_1$ and $\alpha_2$ representing the pathloss exponent in LoS and NLoS  links, respectively.
For a greater reality match, one may increase the number
$K$ of segments to account for extra degrees of
obstruction severity such as concrete building vs. wood-walled structure vs.
foliage etc.
Of course, more segment values means added complexity as
well as a greater noise sensitivity when doing model classification.



\subsection{Map-aided vs. probabilistic attenuation models} \label{sec:pb_ch_model}
The segmented pathloss model is highly dependent on the relative UAV-user locations and the local terrain.  
In map-aided approaches, the segment value
$s$ is directly predicted from a 3D terrain map. For example, if the straight line connecting UAV and user location is obstructed by a blockage or building, we can infer that the radio link is of NLoS  nature. 
In the absence of maps,  {\em probabilistic} segmented models are used, whereby
the segment value is simply governed by a likelihood parameter. In the popular two-segment model with LoS and NLoS segments, the LoS probability $p_{LoS}(\theta_n)$ is used to obtain this likelihood. The LoS probability $p_{LoS}(\theta_n)$ is a logistic type of function applied on $\theta_n$ where $\theta_n$ is the elevation angle between the $n$-th user and the UAV. The intuition is that if the elevation is close to 90 degrees, the probability of LoS will approach 1, while if the elevation is small, a high probability of blockage is expected. Such probabilistic path loss models are global by nature in the sense that the model parameters are trained once and are the same across all ground nodes \cite{HouSitLar:J14}.
While this approach can give useful
insights into the role played by certain system design parameters, it often falls short in real-life robotic placement problems because of the lack of performance guarantees.
Instead, for an actual implementation, it is essential to exploit local information relevant to
the terrain surrounding the ground nodes and the robot. There are several ways to
infer such map-based information in practice, be it from 3D terrain data or
radio measurements from the scene of interest as we  review next.

\section{Learning Maps for Flying Radios}

\label{Sec:Learningmaps} 

\begin{figure}[t!]
\centering \includegraphics[width=1\columnwidth]{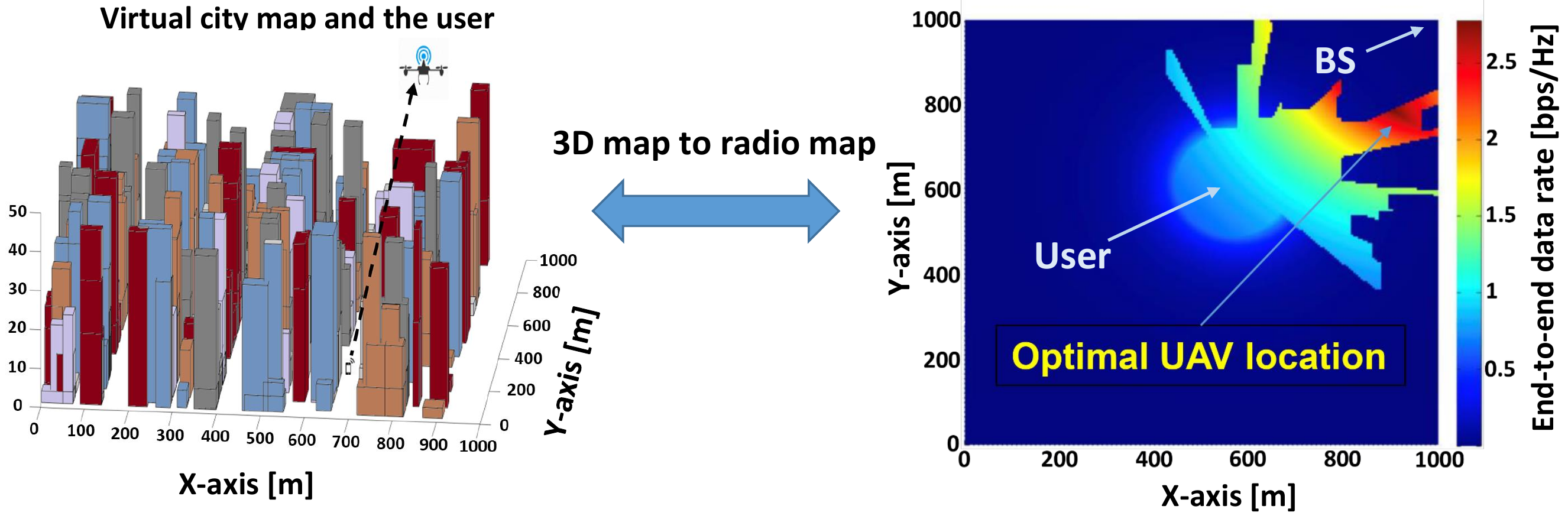} \caption[]{ The end-to-end user-relay-BS throughput map (right figure) for different UAV positions with fixed height over a given city (left figure). The optimal location for the UAV is where it can maintain LoS links to both the user and the BS.}
\label{maps}
\end{figure}

Various kinds of maps
can be used to predict the channel quality 
between a UAV and a ground user. These include 3D terrain or building maps, radio (or RSSI) maps,
and (system-level) throughput maps.  Radio and 3D maps are strongly inter-related via the existence of a reliable path-loss model, as illustrated in Fig. \ref{maps}. This gives rise to powerful {\it joint} radio and 3D map estimation opportunities, which are later mentioned.  Generally,
radio and  throughput  maps are 5D objects as the RSSI or throughput is
specified for each pair of 3D UAV location and 2D ground user location\footnote{In this paper we mostly consider outdoor service. If indoor users
are accounted for, a 3D model for user locations ensues.}. The throughput map may include the effect of finite backhaul 
between the fixed infrastructure and a UAV acting as a relay. A simplifying
relay model is based on the decode-and-forward framework whereby the capacity
is governed by the minimum capacity between the BS-to-UAV link and UAV-to-user
link throughput, respectively.  Note that the BS-to-UAV link
rate is more easily predictable from a distance-based path loss model
as the UAV is often assumed to maintain LoS with the
BS. This leaves the system designer with the sole (yet challenging)
task of predicting the actual link strength between the UAV and the
ground user. Once maps are obtained, it becomes possible
to place the UAV at a throughput-maximizing location or path.

The advantage of map-based placement over a probabilistic approach  is well illustrated in Fig. \ref{maps}
which shows the link throughput for a UAV-relay enabled
communication link between a fixed BS and one ground user.
While a probabilistic approach would always place the UAV somewhere
on the axis between the BS and the ground user so as to minimize path
travel length, the throughput map predicts an optimal UAV location well 
off the BS-user axis. Generally, map-based placement determines the best trade-off
between minimizing distances (i.e., staying close to the BS-user axis)
and exploring LoS opportunities at  other locations to
boost user signal power. 
 

\subsection{Learning radio maps from sparse UAV-borne RSSI measurements}
Radio or 3D maps
are not always readily available and, as flying time is costly,  often
need to be reconstructed from a limited set of radio measurements.
Let us consider the problem of predicting the air-to-ground link
quality at arbitrary UAV locations for one fixed ground
user location. The problem of radio map reconstruction in the fixed cellular context is not new. In fact recent advances using deep learning have been reported such as in \cite{levie2021radiounet}.
In our context, it is assumed that a finite number of RSSI measurements
have been collected beforehand by the UAV for training. Typically, the training data points offer a sparse representation
of the radio map, which must then be reconstructed at all locations
where measurements are missing. Note that this can be done using standard
or adapted machine learning techniques. Generally speaking, the map
reconstruction may be {\em direct} or {\em model-based} 
\cite{CheEsrGesMit:C17}. For a direct reconstruction,
one relies on the hidden correlation structure between RSSI levels
at neighboring training data points and a new candidate UAV location
in order to generate an estimate of the RSSI at the new location.
Averaging of RSSI levels has been considered in 
\cite{RomKimGiaLop:J16} using kernel regression methods. Alternatively,
a model-based approach can be used which exploits radio expert knowledge
in the form of the segmented channel models shown earlier. We need to first estimate the unknown model parameters
from the training data. Unfortunately, it is not known a priori to which segment each of the training
data points actually belongs to. To circumvent the absence of segment
labels in the data, a {\em joint} classification and parameter
estimation approach is proposed. The basis of this approach
is two-stage iterative process where classification of the data points
is first carried out by way of a clustering algorithm on the basis
of pre-estimated model parameters. Then the propagation parameters for each segment can be re-estimated
using maximum-likelihood estimation or least-squares estimation on
each corresponding subset of the data \cite{CheEsrGesMit:C17}. 

\begin{figure}[t!]
\centering \includegraphics[width=1\columnwidth]{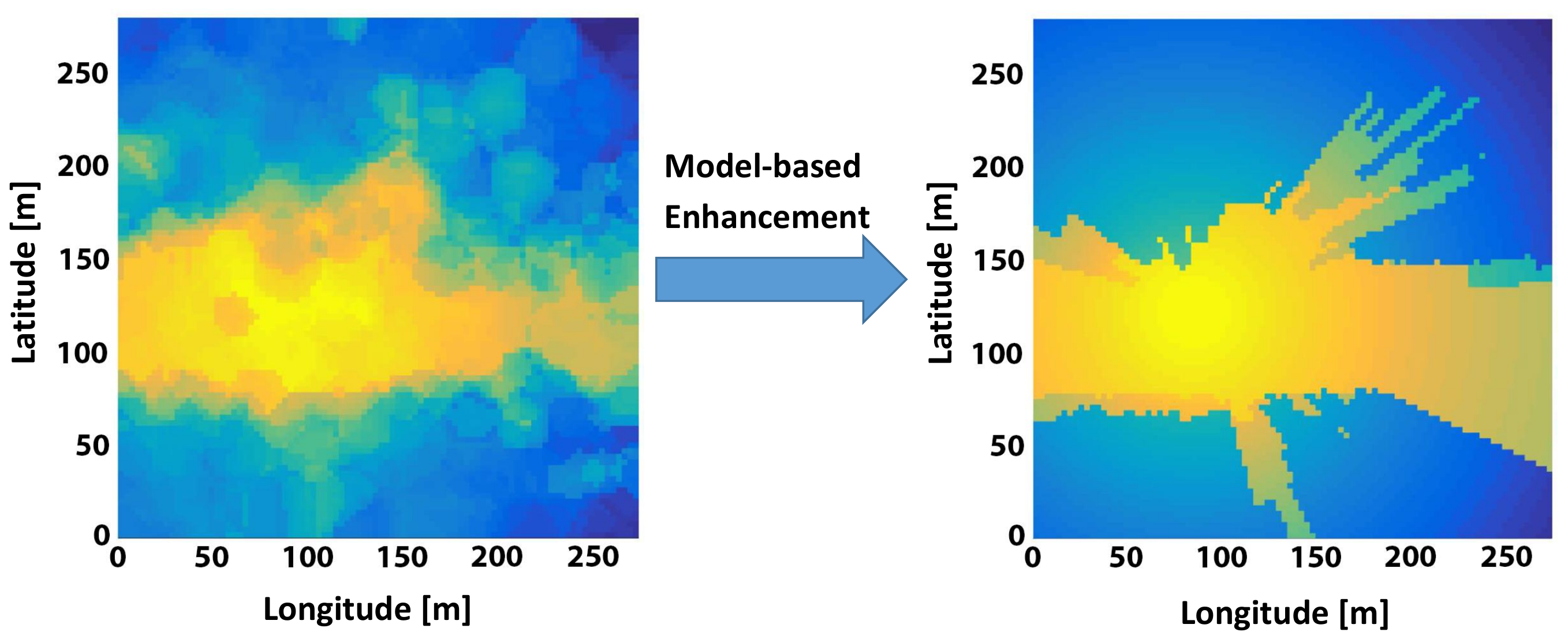} \caption[]{Visual performance for map reconstruction: In contrast with the direct KNN-based reconstruction  approach (left), the exploitation of expert channel model knowledge allows us for crisp reconstruction of the various segments (right).}
\label{reconstruction}
\end{figure}

In Fig. \ref{reconstruction}, the performance of model-based radio
map reconstruction is illustrated and compared with a direct reconstruction
approach in the RSSI domain using the kernel trick 
\cite{RomKimGiaLop:J16} for a selected area at center Washington
DC \cite{CheEsrGesMit:C17}. The UAV moves at a fixed height 50 meters
above ground to visit 400 randomly selected training locations to
measure the link quality to 100 random ground user locations. The model-based
approach can reconstruct the crisp shapes of the propagation segments with
accurate prediction on the channel gain.

\subsection{Radio maps vs. 3D terrain maps}

Note that the above radio map reconstruction is carried out for one
particular fixed ground user location. However, in the presence of
$N>1$ ground users, a separate radio map per user must be estimated
in principle such that an overall system-level throughput map can
be drawn\footnote{ Multi-user throughput may be defined in a number of ways, including
sum throughput, fair throughput, worst case throughput etc. In all
such cases we assume orthogonal multiple-access however.}. Note that the throughput map could in principle be reconstructed directly from
sparse measurements done directly in the throughput domain, hence
bypassing the radio map altogether. However this approach does not leverage
expert knowledge related to channel models. On the other hand, it is desirable to exploit the
inherent correlation existing between radio conditions for ground
users that are close to each other. A powerful way to exploit blockage correlation
across users is to introduce the 3D terrain map as auxiliary information.
The use of 3D terrain maps can be explicit or implicit. If a 3D city map is available, link strengths are directly
predicted from it using ray tracing followed by classification
into one of the $K$ model segments.

\paragraph{Joint radio and 3D map reconstruction}

In some cases, the 3D terrain map may not be available. In fact obtaining a 3D map of surrounding may be the actual goal in mind (e.g. sensing). In this case,
 it is still possible to reverse engineer the radio
map back to the 3D terrain map domain. In practice, this is again
done using the UAV's RSSI measurement data set. While jointly estimating
the model parameters, such RSSI levels are classified to one of two
segment values (i.e. LoS or NLoS). In turn, a UAV-user link which
is classified as LoS reveals that any building located between the
two must be lower than the UAV-user axis, yielding a set of inequality constraints.
A picture of the 3D building map emerges when aggregating these equations
through a large enough number of RSSI measurements from scattered
users in the city. It is then possible to exploit the common underlying
structure between the radio map and the 3D terrain map. This is done
by feeding the building height estimates obtained above, back into
the RSSI prediction model so as to have a more complete radio map
estimate. This procedure can be iterated until no more improvement
is achieved. Finally another approach to enhance map reconstruction consists in complementing radio data with additional vision sensor data (camera or lidar) followed by a suitable fusion algorithm. 
 


\section{Map-based path planning of flying access points}

In the above section, methods for acquiring useful map information on the basis of measurements carried out by radio-enabled UAVs were surveyed. We now turn our attention to the design of map-aided optimal paths for UAVs in the DaaR and DaaB contexts.

\label{sec:placement} 

\subsection{Placement of  UAV based on nested propagation segments}
Let's assume a scenario in which ground users performing special tasks in an area where terrestrial
connectivity is limited or degraded, e.g. military patrolling in remote
areas, site or plant inspection by humans or rovers after a natural
disaster etc.
 Such scenarios can benefit from the deployment of a
user-dedicated UAV-mounted relay.  The challenge
is for the drone to self-position at a location allowing to maximize
the throughput of the end-to-end relay channel.  Since the link between the UAV
and BS tower is likely to be relatively clear of obstacles (assuming
high enough BS antenna), the intuition behind optimal UAV position
lies in striking a balance between keeping path travel distances low
(both for the backward link to the BS as well for the forward
link to the users) {\em and} maintaining a good link quality to
the users by discovering LoS opportunities. Although the optimal position
can be, in theory, computed offline  using a global search over the  
radio map,  it is desirable to have a method which only
requires {\em local} exploration of the map. Such benefit is made
possible by exploiting an interesting {\em nested propagation property}.


The property reflects the notion of LoS irreversibility. Assume a
UAV flies at constant height, initially far off and moving towards
a ground node. Also assume the UAV is initially located in the NLoS
region of this node, then it will hit LoS region once the UAV gets
close enough to the ground node. LoS irreversibility predicts that
LoS will  be maintained without interruption until the UAV reaches
the spot right above the node's location. Intriguingly, an implicit condition for this is that the buildings and other (large) obstacles
on the ground have a convex shape, which fortunately is often the case. For a more general segmentation ($K>2$), this property extends by
arguing that the UAV-user channel tends to become less obstructed
as the UAV moves towards the ground user. For a fixed given ground node location, denoting
by ${\cal D}_{i}$ the region formed by UAV locations for which $s=i$,
we have that ${\cal D}_{i}$ is {\em nested} inside ${\cal D}_{i+1}$ (e.g. the LoS region is nested inside the NLoS region).
Note that the nested propagation regions property conveys some useful
structure to radio maps and it can be shown that the optimal UAV position
can {\em only} be either the BS-user axis {\em or} one segment's boundary. 
As a result, it is possible to derive globally convergent algorithms with linear search complexity in terms of the BS-user distance \cite{CheGes:J20}.


\subsection{Intelligent IoT data harvesting} \label{mapc}

When the UAV addresses the connectivity of multiple ground nodes,
it can be shown \cite{ZenZha:J17} that the optimal placement involves designing
a {\em path} allowing the UAV to cycle through points located above,
and in the vicinity of these ground nodes, as opposed to having the
UAV hover above a static location. To what extent that path takes
the UAV near the nodes depends on the limited on-board battery budget.
The problem of designing an optimal communication path can be formulated
as an extension of UAV static placement where the UAV location
is replaced by a vector of time-discretized locations satisfying extra
dynamical constraints (bounded velocity, acceleration and deceleration);
the throughput reflects a summation over the rates offered to the
multiple ground nodes. With the use of probabilistic segmented channel models, the optimal
path design is amenable to classical optimization tools thanks to
the differentiability of the RSSI with respect to UAV location \cite{wu2021comprehensive}.
The disadvantage of this approach is that local terrain features are ignored and the obtained path
cannot offer performance guarantees on communication quality.


\subsection{Map compression}  \label{mapc_compression}
 When the RSSI at a given drone location
is drawn from a 3D or radio map,  we obtain rich local information enabling accurate throughput predictions at the UAV. Unfortunately, the irregular shapes of the LoS
regions in the many user case renders the throughput function non-differentiable
as a function of the UAV location. A key to solving this problem
resides in the idea of map smoothing or {\it compression} \cite{Esra19}.  The goal is to
 preserve essential node-location dependent channel behavior while smoothing out other map details.
In practice, this is done by converting map data into
a reliable \textit{node location dependent} LoS probability model
which is now modeled by $p_{LoS}^{n}(\theta_n)$
where $\theta_n$ denotes
the elevation angle for ground node $n$ and the LoS probability is now made dependent on the location of ground node $n$ by using node-dependent logistic regression parameters.
Such parameters
can be learned  (e.g. \cite{Esra19}) from
a training data set formed by a set of tentative UAV locations around
the $n$-th ground node along with the true LoS status obtained from
the 3D map. Interestingly, the extended model above can be seen as
{\em localized} extension of the classical  probability
model. The key advantage in using the 
local probability model over the {\em global} one
is that it discriminates between the ground nodes in terms of LoS opportunities
they allow, while going around the non differentiability
issues created by  raw map data.

\begin{figure}[t!]
\centering \includegraphics[width=0.95\columnwidth]{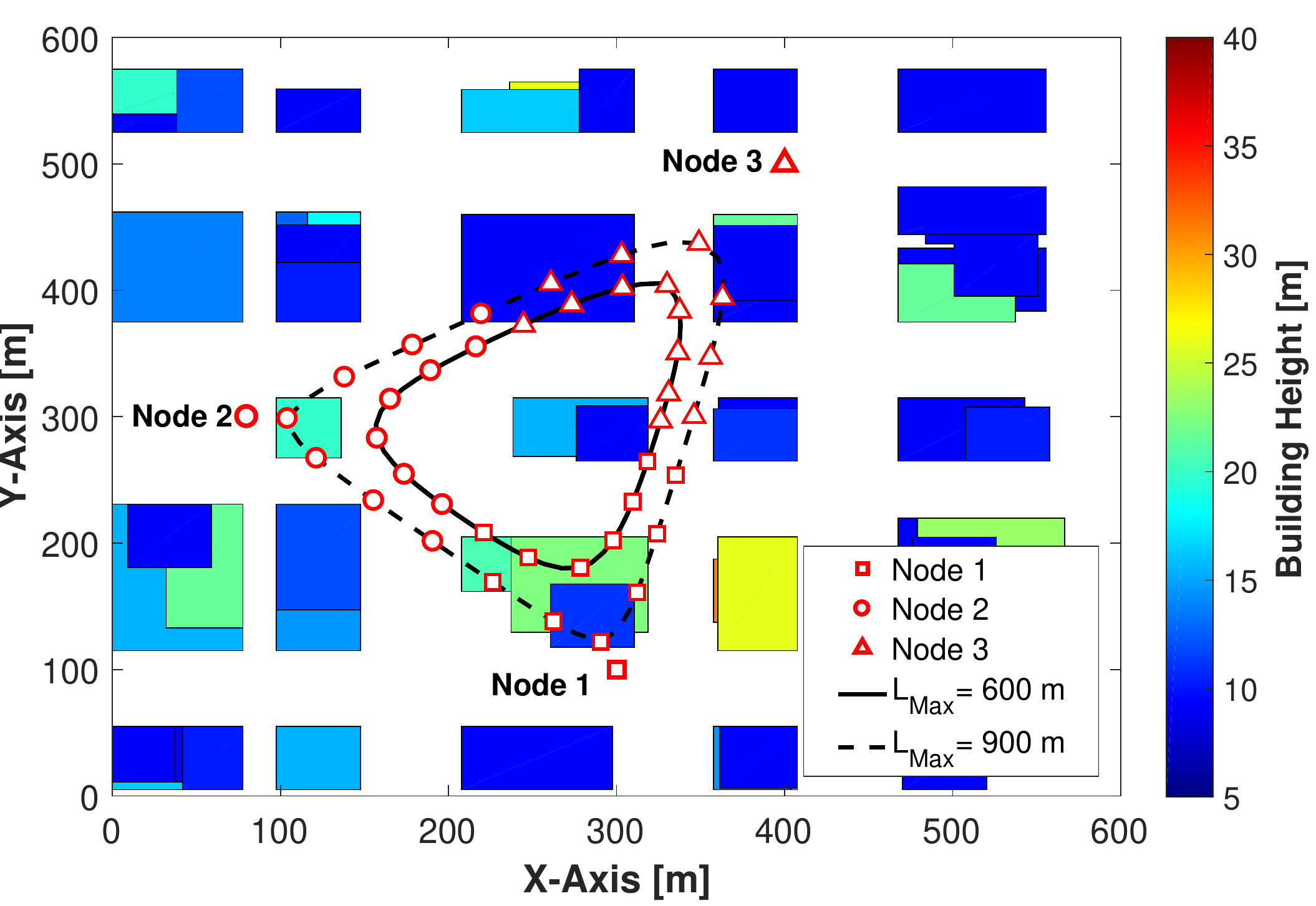} \caption[]{  An example of path planning in an IoT data harvesting setting with 3 ground sensors and for different UAV trajectory lengths ($L_{\text{Max}}$). As the length of the trajectory increases, the UAV moves towards the ground sensors to improve the link quality. The markers on the trajectory indicate the UAV at that time collects data from the node shown with a similar marker.}
\label{iot}
\end{figure}

An example of a path obtained under the map compression approach for
an IoT setting with 3 nodes is shown in Fig. \ref{iot}. The optimal path design exploiting compressed map information
allows the UAV to exploit LoS opportunities when possible, for some
of the ground nodes, while
designing a path under classical probabilistic channel models, tends to treat all ground nodes in a similar
way. As a result of using the map compression technique, the UAV can decide to stay at a
distance from node 3 which is located in a relatively open area while
cycling closer to other nodes which are shadowed away by taller buildings that can bring a considerable gain in the amount of data collected from the nodes \cite{Esra19}. 

Fig. \ref{fig:iot_map_comparison} shows the advantage of using the map compression method over different algorithms for UAV trajectory design in an IoT scenario where a DaaB collects data from 6 ground sensors. In the deterministic algorithm, an optimal trajectory is generated by considering a single deterministic LoS channel model for the links between the UAV and ground sensors. In the probabilistic algorithm, a probabilistic segmented channel model, which was introduced in section \ref{sec:pb_ch_model}, is used \cite{Esra19}.

 


\begin{center}
\begin{figure}[t!]
\centering \includegraphics[width=0.9\columnwidth]{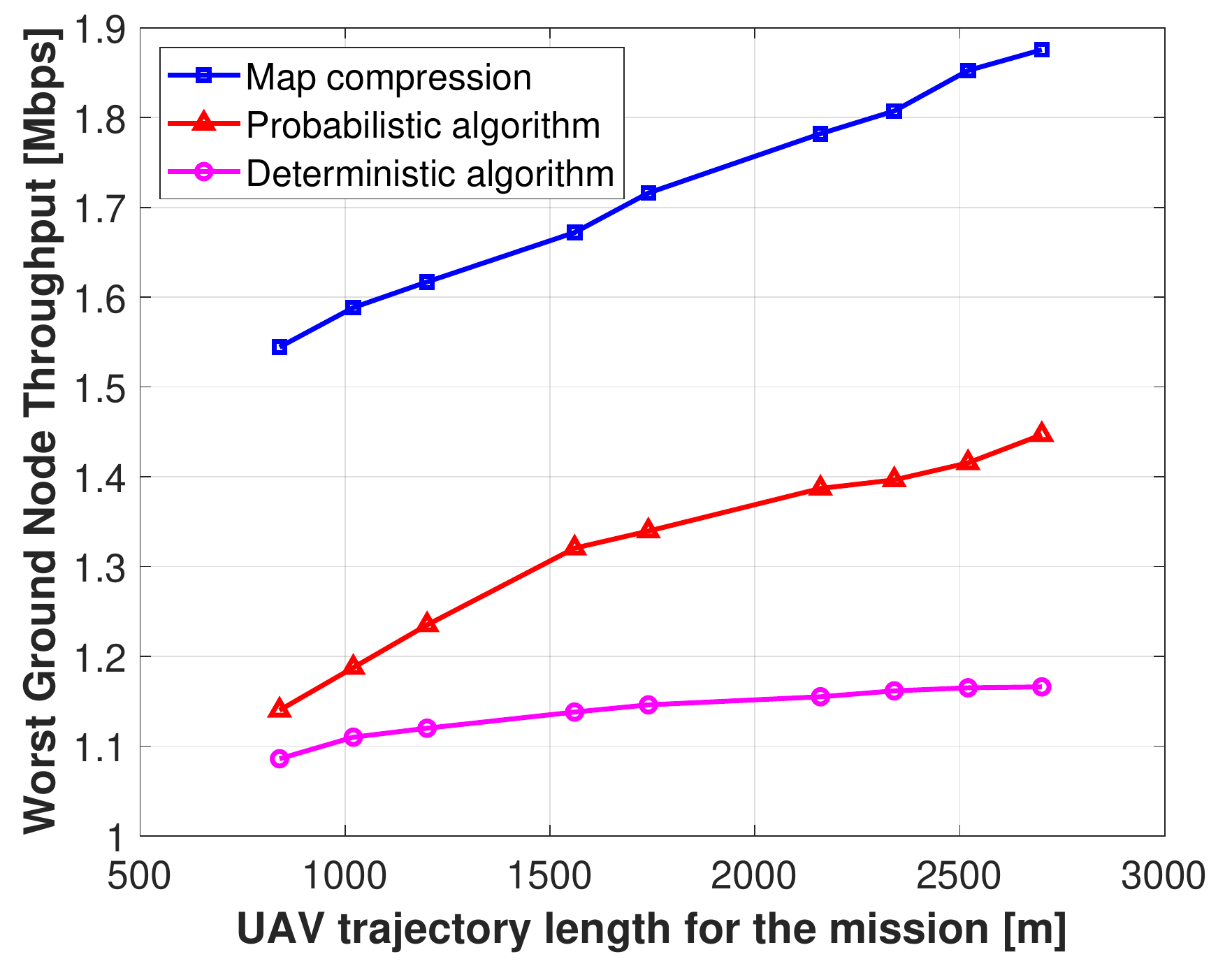} \caption[]{ The performance comparison for the map compression-based UAV trajectory design algorithm versus increasing the trajectory length in an IoT scenario where a DaaB collects data from 6 ground users.}
\label{fig:iot_map_comparison}
\end{figure}
\par\end{center}




\section{Robot-aided RF sensing based on active learning}

So far we have seen the use cases of UAV acting as a flying RAN device aiming to improve the wireless connectivity services towards ground nodes. 
In the context of 6G research where there is a growing convergence between communication and sensing systems \cite{bourdoux20206g}, we can consider the DaaS scenario where the UAV acts as a flying RF sensor that can assist with sensing and localization services. Contrary to static or uncontrolled mobile devices in the network, we can optimize the UAV trajectories to improve the sensing and localization performance. Specifically, UAV path planning problems that constitute a good trajectory to collect the most informative measurements, among all feasible paths, satisfying a duration or energy budget constraint can be formulated. In machine learning and robotics, this problem is sometimes referred to as active learning or optimal experiment design \cite{taylor2021active}.

Interestingly, the map can one more time help to predict a UAV trajectory allowing us to collect maximally informative measurements. For instance, the availability of the 3D map can let us determine where and when the UAV can maintain LoS connections to the users for collecting the measurements which tend to be more suitable for precise sensing (i.e. LoS measurements are subject to less shadowing noise than NLoS measurements). It is shown in \cite{esrafilian20203d} that by exploiting the 3D map an improvement of about $70 \%$ over the other approaches in the user localization accuracy can be obtained.

\section{Prototypes}

The problem of autonomous placement of micro-UAVs as flying radios
(LTE relays or BSs) has been the subject of relatively few
practical deployments and prototypes to date. In fact, much prior work
examining the design of UAV-based RANs is based on the simplifying
idea that the UAV serves as a mechanical flying device on which a BS is mounted, hence communication and navigation functionalities are mostly kept decoupled. Hence, the potential associated with optimized 3D placement with UAV BSs
in theoretical works cannot be fully demonstrated with such prototypes in real-world scenarios.

\begin{figure}[t!]
\centering \includegraphics[width=0.9\columnwidth]{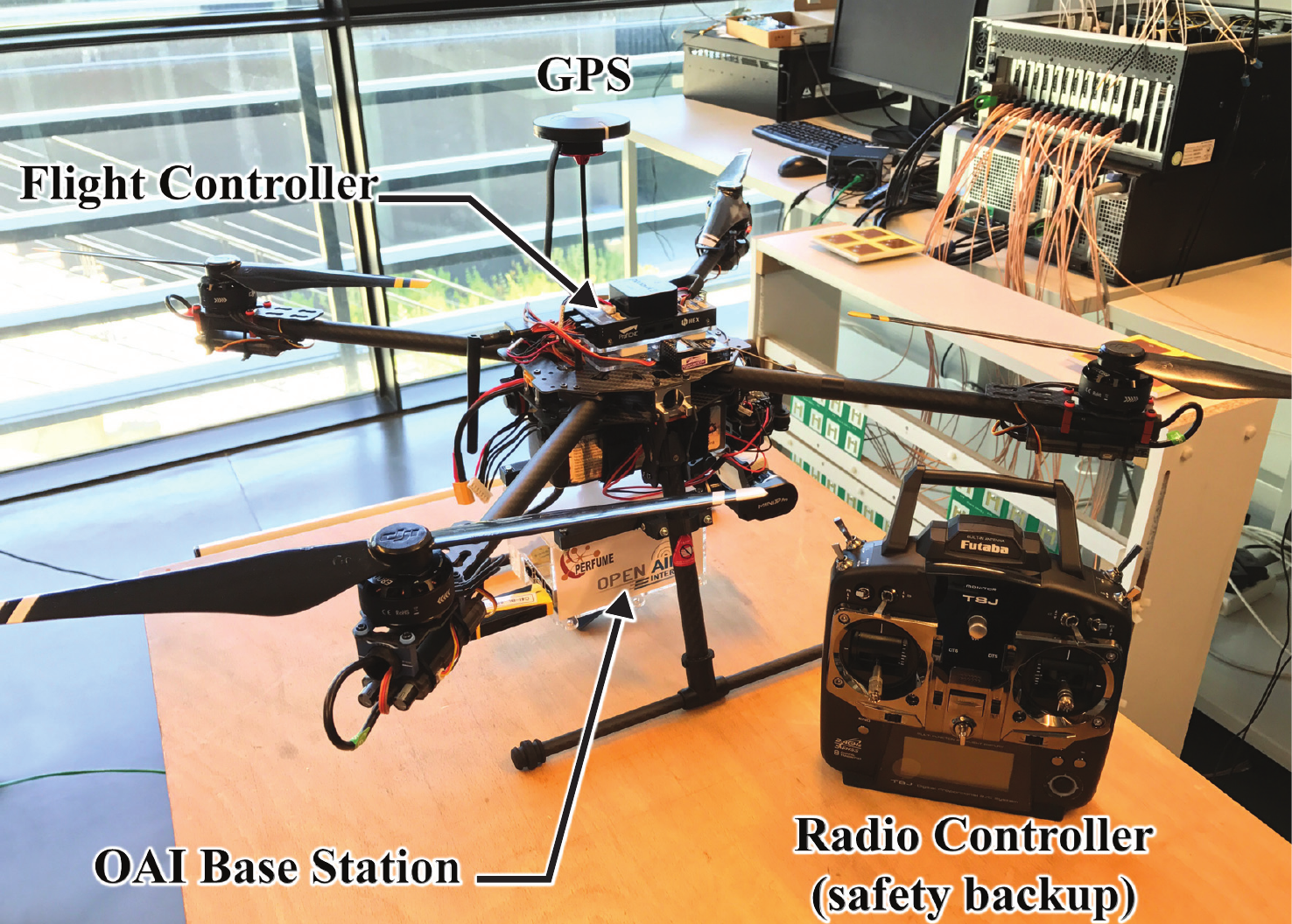} \caption[]{The ReBot prototype: A fully integrated autonomous UAV with end-to-end
LTE relay functionality.}
\label{rebot} 
\end{figure}

In \cite{gangula2018flying}, the 
Rebot (Relay Robot) concept was presented. The Rebot functions both
as an outdoor LTE relay between ground users and a fixed BS, as well
as a fully customized autonomous robot capable of positioning itself at a throughput
maximizing location. The Rebot's communication layer embedded
on the UAV is based on the OpenAirInterface (OAI), which
is an open-source reference implementation of 3GPP standards
running onboard the UAV using commodity Linux computing equipment.
In \cite{gangula2018flying}, a video recording
of the experiment on the EURECOM campus is also captured, illustrating 
the throughput advantage and the machine learning-driven self-placement
and tracking capabilities of the Rebot. The UAV continuously collects and processes radio measurements
over the flight path to update the estimate for the optimal placement solution. Different parts of the UAV are shown in Fig. \ref{rebot}. 

\section{Perspectives} \label{persp}

The deployment of UAV-aided wireless networks offers a host of mixed robotic-communications analysis problems \cite{wu2021comprehensive}. {\em Real-time} placement algorithms remains a central issue  for which the use of machine learning-driven map-aided methods seem promising. Interestingly, the useful interactions between 3D mapping and UAV path planning have also been recently investigated in scenarios beyond the sole UAV-aided communication use cases. For instance the role of maps was highlighted in the context of UAV positioning for optimal wireless power transfer \cite{MoHuangXu:J19}. In the DaaT scenario, a central issue is the design of safe UAV paths that allow the robot to reach a prescribed destination while satisfying connectivity (from network) constraints all along the way. The use of radio maps was shown to be highly beneficial also in that context \cite{zhang2020radio} as they enable more accurate connectivity predictions than probabilistic channel models. In such DaaT scenario as well, maps create an information richness dilemma which can be mitigated using the compression method surveyed in Sec. \ref{mapc_compression}.

Moreover, the ideas and tools discussed in the above sections can be extended to multi-UAV scenarios where there are several UAVs (i.e. a swarm of UAVs) cooperating to provide better coverage and services to users. However, when it comes to the multiple UAVs, we still need to face the same problem as in the single UAV case, because when a UAV in the swarm communicates with a ground node, the communication link may probably be blocked by obstacles. Therefore, the exploitation of the map and ideas proposed in this paper can still be beneficial for the multi-UAV case to guarantee the link quality between UAVs and ground users.
 
\section{Acknowledgment}

Part of the work by David Gesbert, Omid Esrafilian, and Rajeev Gangula was funded via the HUAWEI France supported Chair on Future Wireless Networks at EURECOM. The part done by Junting Chen was supported by the National Key R\&D Program of China with grant No. 2018YFB1800800, by National Science Foundation of China No. 92067202 and No. 62171398, by the Shenzhen Science and Technology Program under Grant No. JCYJ20210324134612033, and by the Guangdong Provincial Key Laboratory of Future Networks of Intelligence. The part done by Urbashi Mitra was supported by ONR N00014-15-1-2550, NSF CCF-1817200, ARO W911NF1910269, Cisco Foundation 1980393, DOE DE-SC0021417, Swedish Research Council 2018-04359, and NSF CCF-2008927.


\bibliographystyle{IEEEtran}
\bibliography{IEEEabrv,Bib_wcm18}


\begin{IEEEbiographynophoto}{David Gesbert} 
[Fellow, IEEE] is Director of EURECOM, France. He received his Ph.D. from TelecomParis, in 1997. Prior to EURECOM, he was with ISL, Stanford University, was a founding engineer of Iospan Wireless Inc., a Stanford spin-off pioneering MIMO-OFDM, and with the University of Oslo. He published about 350 articles and 25 patents, 7 of them winning IEEE Best paper awards, and was named a Highly Cited Researcher in computer science.  He is a Board Member of the OpenAirInterface Software Alliance. He was an awardee of the ERC Advanced Grant and the Grand Prix in Research from IMT and the French Academy of Sciences.
\end{IEEEbiographynophoto}


\begin{IEEEbiographynophoto}{Omid Esrafilian} 
[Member, IEEE] received his Ph.D. degree from Sorbonne University and EURECOM, France, in 2020. He is currently working at EURECOM as a
research engineer building prototypes for autonomous aerial cellular relay drones capable of providing flexible and enhanced (LTE, 5G) connectivity to mobile users. He was the recipient of several awards in the national and international robotics competitions as well as the award for "Fundamental research project of the year", 2019, by the French SCS Research and Industry Cluster. His main research interests include unmanned aerial vehicle communication systems, machine learning, and robotics.
\end{IEEEbiographynophoto}


\begin{IEEEbiographynophoto}{Junting Chen} 
[Member, IEEE] received the B.Sc. degree in electronic engineering from Nanjing University, Nanjing, China, in 2009, and the Ph.D. degree in electronic and computer engineering from The Hong Kong University of Science and Technology (HKUST), Hong Kong, SAR, China, in 2015. He is currently an Assistant Professor with the School of Science and Engineering and the Future Network of Intelligence Institute (FNii), The Chinese University of Hong Kong, Shenzhen (CUHK-Shenzhen), China. He works on radio map learning and applications, UAV assisted communications, and, more generally, machine learning and optimization for wireless communications and localization. 
\end{IEEEbiographynophoto}


\begin{IEEEbiographynophoto}{Rajeev Gangula} [Member, IEEE]
obtained his Ph.D. degrees from Telecom ParisTech
(Eurecom), France, in 2015. He is currently working at Eurecom as a
research engineer building prototypes for autonomous aerial cellular
relay drones capable of providing flexible and enhanced (LTE, 5G)
connectivity to mobile users. His research interests lie in the areas of
optimization, communication theory and development of unmanned aerial
vehicle communication system.
\end{IEEEbiographynophoto}


\begin{IEEEbiographynophoto}{Urbashi Mitra } [Fellow, IEEE] is currently the Gordon S. Marshall Professor in Engineering at the University of Southern California. She is the recipient of: the 2021 USC Viterbi Senior Research Award, 2017 IEEE Women in Communications Engineering Technical Achievement Award, a 2015 UK Royal Academy of Engineering Distinguished Visiting Professorship, a 2015 US Fulbright Scholar Award, a 2015-2016 UK Leverhulme Trust Visiting Professorship, IEEE Communications Society Distinguished Lecturer, 2012 Globecom Signal Processing for Communications Symposium Best Paper Award, 2012 US National Academy of Engineering Lillian Gilbreth Lectureship, the 2009 DCOSS Applications \& Systems Best Paper Award, 2001 Okawa Foundation Award, 2000 Ohio State University’s College of Engineering Lumley Award for Research, and a 1996 National Science Foundation CAREER Award.
\end{IEEEbiographynophoto}

\end{document}